\documentclass{aa}  
\usepackage{color}
\usepackage{graphicx}
\usepackage{lscape}
\usepackage{natbib}
\usepackage{natbib,twoopt}
\usepackage[varg]{txfonts}
\usepackage{url}
\usepackage{xspace}
\usepackage{amssymb}
\usepackage{stmaryrd}
\usepackage{siunitx}
\usepackage{textcomp}
\usepackage{placeins}
\usepackage{float}

\bibpunct{(}{)}{;}{a}{}{,}

\def\etal{{et\,al.}\ }

\newcommand{\gaiaf}{\object{Gaia DR3 5253439589461750912}\xspace}
\newcommand{\gaiafs}{\object{Gaia DR3\,52}\xspace}
\newcommand{\gsc}{\object{GSC\,08265$-$00355}\xspace}
\newcommand{\gscs}{\object{GSC\,08265}\xspace}
\newcommand{\ucac}{\object{UCAC4 108$-$030787}\xspace}
\newcommand{\ucacs}{\object{UCAC4\,108}\xspace}

\newcounter{Rco}
\newcommand{\Ionst}[1]{\setcounter{Rco}{#1}\Roman{Rco}}
\newcommand{\Ion}[2]{\mbox{#1\,{\scriptsize\Ionst{#2}}}}

\newcommand{\logg}{\mbox{$\log g$}\xspace}

\newcommand{\Teff}{\mbox{$T_\mathrm{eff}$}\xspace}

\newcommand{\ebv}{$E_\mathrm{B-V}$\xspace}

\newcommand{\vrad}{$v_\mathrm{rad}$\xspace}

\newcommand{\Lsol}{$L_\odot$}
\newcommand{\Msol}{$M_\odot$}
\newcommand{\Rsol}{$R_\odot$}

\begin{document}

\title{Three new hot hydrogen-deficient pre-white dwarfs
}

\author{Klaus Werner\inst{1} \and Nicole Reindl\inst{2} \and
  Max Pritzkuleit\inst{3} \and Stephan Geier\inst{3}}

\institute{Institut f\"ur Astronomie und Astrophysik, Kepler Center for
  Astro and Particle Physics, Eberhard Karls Universit\"at, Sand~1, 72076
  T\"ubingen, Germany\\ \email{werner@astro.uni-tuebingen.de} 
\and
  Landessternwarte Heidelberg, Zentrum f\"ur Astronomie, Ruprecht-Karls-Universit\"at, Königstuhl~12, 69117 Heidelberg, Germany  
\and
  Institut f\"ur Physik und Astronomie, Universit\"at Potsdam, Karl-Liebknecht-Stra\ss e 24/25, 14476, Potsdam, Germany
}

\date{Received 28 November 2024 / Accepted 18 December 2024}

\authorrunning{K. Werner \etal}
\titlerunning{Three new hot hydrogen-deficient pre-white dwarfs}

\abstract{We have detected three new hydrogen-deficient (H $<$ 0.001 mass fraction) pre-white dwarfs (WDs) with helium-dominated atmospheres. The first object is a relatively cool PG1159 star (effective temperature \Teff = 72\,000\,K) that has the lowest surface gravity of any PG1159 star known (\logg = 4.8). It is a PG1159 star in the earliest pre-WD phase. The second object is a hot subdwarf O (sdO) star (\Teff = 50\,000\,K, \logg = 5.3) with high carbon and oxygen abundances. It is only the third known member of the recently established CO-sdO spectral class, which comprises stars that are thought to be formed by a merger of a disrupted low-mass CO WD with a higher-mass He WD. The third object is one of the rare stars of spectral type O(He) (\Teff = 90\,000\,K, \logg = 5.5).}


\maketitle
%

\begin{table*}[t]
\tiny
\begin{center}
\caption{
Program stars analyzed in this paper. 
  \tablefootmark{a}
}
\label{tab:stars} 
\begin{tabular}{ccccccc}
\hline 
\hline 
\noalign{\smallskip}
Full name & Short       name & $\alpha$&$\delta$& G-Mag  & Obs.Date   & Exposure \\ 
          &                  &         &        &        & YYYY-MM-DD & [s]      \\    \hline
\noalign{\smallskip} 
\gaiaf & \gaiafs & 10 11 10.33 & $-$60 40 36.67 & 14.519 & 2023-02-12 & 1080 \\
\gsc   & \gscs   & 13 35 27.76 & $-$48 16 19.30 & 12.456 & 2022-12-19 & 900  \\
\ucac  & \ucacs  & 10 29 26.52 & $-$68 33 21.47 & 13.785 & 2023-02-12 & 1080 \\
\noalign{\smallskip}
\hline
\end{tabular} 
\tablefoot{  \tablefoottext{a}{Coordinates and G magnitudes from \emph{Gaia} DR3
    \citep{2022yCat.1355....0G}.  }}
\end{center}
\end{table*}

\section{Introduction}
\label{sect:intro}

Hot hydrogen-deficient stars approaching the white dwarf (WD) cooling sequence are the result of noncanonical evolution. They either form as single stars after violent interior processes or are the outcome of close binary evolution. There is a rich variety of these hot pre-WDs in terms of the chemical composition of their atmospheres and their position in the Hertzsprung-Russell diagram. They are assigned to different spectral classes, some of which have only a few members. In order to clarify their evolutionary history, more need to be found. In this paper we analyze three new H-deficient stars that belong to three different spectral classes. 

We found one new O(He) star. To date, there are only 13 known stars of this class \citep[for the most recent discoveries, see][]{Jeffery2023}. They have helium-dominated atmospheres with effective temperatures in the range \Teff = 80\,000 -- 200\,000\,K and surface gravities of \logg = 5.0 -- 6.7. Some of them have residual hydrogen and traces of C, N, O, and Si. Different abundance patterns point at different evolutionary scenarios.  The O(He) stars that are not associated with a planetary nebula (PN) are likely He WD mergers, while the evolutionary status of O(He) central stars with a PN is less clear (see the discussion in \citealt{2014A&A...566A.116R}). 

In addition, we discovered a hot helium-rich subdwarf O star (sdO) with unusually high carbon and oxygen abundances. It is only the third known member of the recently established group of CO-sdOs \citep{2022MNRAS.511L..66W}. They have temperatures and gravities around \Teff = 50\,000\,K and \logg = 5.5 and C and O abundances of about 0.2. (All abundances in this paper are given in mass fractions.) They probably were formed by a merger of a CO WD and a He WD \citep{2022MNRAS.511L..60M}. It was speculated that these objects could evolve into PG1159 stars.

There are approximately 70 known PG1159 stars, and they cover a wide range in the Kiel diagram (\Teff = 60\,000 -- 250\,000\,K and \logg = 5.3 -- 8.3; see, e.g., \citealt{2006PASP..118..183W} and for the latest discoveries \citealt{2024A&A...686A..29W}). Their atmospheres are composed of He, C, and O in rather different amounts, ranging between He-dominated atmospheres (He $>0.9$) and He-deficient atmospheres dominated by C and O. The majority of PG1159 stars are thought to be the result of a late or very late thermal pulse (LTP or VLTP), but a small fraction could stem from the abovementioned CO-sdOs or from the surviving components of double-degenerate supernova Ia explosions \citep{2024A&A...689L...6W}. In this paper we investigate a new star classified as a PG1159  but with unusual characteristics.

The remainder of this paper is organized as follows. In Sect.\,\ref{sect:analysis} we present the observations and the spectral analysis of the stars, and we derive their physical parameters in Sect.\,\ref{sect:parameters}. The results are summarized and discussed in Sect.\,\ref{sect:discussion}.

\section{Observations and spectral analysis}
\label{sect:analysis}

Spectra of our three program stars were obtained with the X-shooter instrument at ESO’s Very Large Telescope (ProgID: 110.23TL; PI: M. Pritzkuleit). They cover the wavelength range 3000 -- 10\,000\,\AA\ with a resolving power of $R\approx10\,000$. The normalized spectra are displayed in Figs.\,\ref{fig:gaiaf} -- \ref{fig:ucac}. Coordinates, magnitudes, and details of our observations of the stars are given in Table~\ref{tab:stars}.

For the spectral analysis we used the T\"ubingen Model-Atmosphere Package (TMAP\footnote{\url{http://astro.uni-tuebingen.de/~TMAP}}) to compute nonlocal thermodynamic equilibrium, plane-parallel, line-blanketed atmosphere models in radiative and hydrostatic equilibrium \citep{1999JCoAM.109...65W,2003ASPC..288...31W,tmap2012}. We computed models of the type introduced in detail by \cite{werner2014} and \cite{2014A&A...569A..99W}. They were tailored to investigate the optical spectra of O(He) stars and relatively cool PG1159 stars. In essence, they consist of the main atmospheric constituents, namely helium, carbon, and oxygen. Hydrogen, nitrogen, and silicon were included as trace elements in subsequent line-formation iterations, meaning the atmospheric structure was kept fixed. 

The model calculations were rather difficult because of numerical instabilities. Convergence of models was only achieved with careful treatment, and it was therefore impossible to compute model grids to determine the atmospheric parameters. For each star, we computed a series of models in order to find a good fit to the observed line spectra. 

\subsection{\gaiaf: An O(He) star}

\gaiaf (hereafter \gaiafs) was classified as a hot subdwarf in the catalog of \cite{2022A&A...662A..40C}. The spectrum of this star is shown in Fig.\,\ref{fig:gaiaf}. It is dominated by \Ion{He}{2} lines. Metal lines are scarce and stem from nitrogen and silicon. Notably absent are features from carbon and oxygen. In detail, we see very weak lines from \Ion{N}{4}, namely from multiplets around 3480\,\AA\ (in absorption) and 5200\,\AA\ (in emission), as well at two singlet lines at 4050\,\AA\ (in absorption) and 6380\,\AA\ (in emission). From \Ion{N}{5} we observe a prominent doublet at 4604/4620\,\AA\, in absorption and the 4945\,\AA\ multiplet in emission. Silicon is detected by the \Ion{Si}{4} 4089/4112\,\AA\ doublet that appears in emission. A very weak \Ion{Si}{4} doublet in absorption is probably also present  at 6668/6701\,\AA. The strongest carbon lines predicted by models are \Ion{C}{4} 4648\,\AA\ and the doublet at 5802/5812\,\AA. Their absence in the observation was used to derive an upper abundance limit. For oxygen, the strongest predicted line is \Ion{O}{5} 5112\,\AA. Its absence from the observed spectrum provides an upper limit for the element abundance.

We constrained the effective temperature and surface gravity by comparing the \Ion{He}{2} line profiles of different models. In addition, the absence of neutral helium lines gives us a strict lower limit for the temperature. At \Teff $< 90\,000$\,K (in the relevant gravity range), we would see a \Ion{He}{1} 5876\,\AA\ absorption line, and \Ion{He}{1} 6671\,\AA\ would appear as an emission line. At \Teff $\geq 100\,000$\,K, the \Ion{N}{5} 4602/4620\,\AA\ doublet would turn into emission. At \Teff = 90\,000\,K, the core of the modeled \Ion{He}{2} 6560\,\AA\ line would have a weak emission reversal, which we indeed observe. At higher and lower temperatures, the emission would be stronger than what we observe or absent, respectively. A good fit to the line profiles of the higher members of the \Ion{He}{2} Pickering series is achieved by models with \logg = 5.5. The line cores of the lower series members, however, are slightly too weak at this value of gravity. A lower value of \logg = 5.3 would fit their line core depths, but then cores of the higher series members become too deep. A good compromise for temperature and gravity is achieved at \Teff $= 90\,000\pm10\,000$\,K and \logg $= 5.5\pm0.2$. In accordance, the relative strength of the \Ion{N}{4} and \Ion{N}{5} lines is well matched by the model with these parameters.

As to the element abundances, we derived an upper limit for hydrogen of H $<$ 0.001. At higher abundances, the models predict a H$\alpha$ emission line similar to the adjacent \Ion{He}{2} emission reversal; however, this is not observed. Our derived upper abundance limits for carbon and oxygen are C $<3\times10^{-5}$ and O $<5\times10^{-4}$. The nitrogen abundance was measured from \Ion{N}{5} 4602/4612\,\AA,\ and we find N = $(3\pm2)\times10^{-4}$. The emission strength of the \Ion{N}{5} 4945\,\AA\ feature does not match this value; however, this problem is probably connected to the unknown wavelength splitting of this multiplet \citep{Werner2022}. The very weak \Ion{N}{4} absorption and emission lines are matched by the model.

The atmospheric parameters of \gaiafs are summarized in Table~\ref{tab:resultsall}, and the respective best-fit model spectrum (with carbon and oxygen abundances set to their upper abundance limits) is displayed in Fig.\,\ref{fig:gaiaf}.

\begin{table*}[t]

\tiny

\begin{center}
\caption{Atmospheric properties and derived stellar parameters of the analyzed stars.
\tablefootmark{a} }
\label{tab:resultsall}
\begin{tabular}{ccccc}
\hline 
\hline 
\noalign{\smallskip}
Parameter                 & \gaiafs              & \gscs                & \ucacs               &           Sun \\ 
\hline 
\noalign{\smallskip}
spectral type             & O(He)                & PG1159               & CO-sdO               &\\
\noalign{\smallskip}
\Teff/\,K                 & $90\,000 \pm 5000$   & $72\,000 \pm 3000$   & $50\,000 \pm 3000$   &\\
\noalign{\smallskip}
$\log$($g$\,/\,cm\,s$^{-2}$) & $5.5 \pm 0.2$     & $4.8 \pm 0.2$        & $5.3 \pm 0.2$        &\\
\noalign{\smallskip}
H                         & $<0.001$             & $<0.001$             & $<0.001$             & 0.74 \\             
\noalign{\smallskip}
He                        & 0.99                &$0.62^{+0.10}_{-0.10}$& $0.85\pm0.05$       & 0.25 \\   
\noalign{\smallskip}
C                         & $<3.0 \times 10^{-5}$& $0.15\pm0.05$        & $0.11\pm0.03$        & $2.4 \times 10^{-3}$\\ 
\noalign{\smallskip}
N                     &$0.003\pm0.002$     &$0.003^{+0.007}_{-0.002}$& $0.005\pm0.003$     & $6.9 \times 10^{-4}$\\ 
\noalign{\smallskip}
O                     & $<5.0 \times 10^{-4}$& $0.23\pm0.05$        & $0.03\pm0.02$        & $5.7 \times 10^{-3}$\\ 
\noalign{\smallskip}
Si                    &$2\pm1 \times 10^{-4}$& $<0.001$             &$(3^{+3}_{-1}) \times 10^{-4}$& $6.6 \times 10^{-4}$\\ 
\noalign{\smallskip}
\vrad\ / km\,s$^{-1}$      & $0\pm 5$            & $-10\pm 5$          & $-15\pm 5$          & \\
\noalign{\smallskip}
$\log$($L$\,/\,\Lsol)      &$3.48^{+0.37}_{-0.29}$&$3.75^{+0.37}_{-0.29}$& $\approx 2.79$      &\\
\noalign{\smallskip}
$\log$($L_\mathrm{SED}$\,/\,\Lsol)&$3.42^{+0.55}_{-0.25}$&$3.54^{+0.18}_{-0.17}$ &  $2.71^{+0.18}_{-0.17}$& \\
\noalign{\smallskip}
$R$\,/\,\Rsol              &$0.23^{+0.08}_{-0.05}$&$0.48^{+0.19}_{-0.10}$& $\approx 0.33$     &\\
\noalign{\smallskip}
$R_\mathrm{SED}$\,/\,\Rsol &  $0.21^{+0.03}_{-0.02}$  & $0.38^{+0.05}_{-0.04}$   &     $0.30^{+0.03}_{-0.02}$  & \\
\noalign{\smallskip}
$M_\mathrm{g}$\,/\,\Msol   &  $0.51^{+0.55}_{-0.25}$  &  $0.33^{+0.34}_{-0.16}$  &  $0.66^{+0.60}_{-0.30}$     & \\
\noalign{\smallskip}
$M$\,/\,\Msol\ (VLTP)      &  $0.52^{+0.02}_{-0.02}$  &  $0.53^{+0.13}_{-0.02}$  & --  \\
\noalign{\smallskip}
$M$\,/\,\Msol\ (merger)    &  $0.65^{+0.06}_{-0.05}$  &  $0.76^{+0.09}_{-0.08}$  & $\approx 0.80$ \\
\noalign{\smallskip}
\ebv\,/ mag                & $0.08\pm0.1$             & $0.08\pm0.1$             & $0.13\pm0.1$ \\
\noalign{\smallskip}
$d$ / pc (\emph{Gaia} parallax)& $2283^{+205}_{-163}$ &$1529^{+115}_{-111}$      & $1765^{+91}_{-72}$    & \\
\noalign{\smallskip}
\hline
\end{tabular} 
\tablefoot{  \tablefoottext{a}{Element abundances given in mass
    fractions. Solar abundances from
    \citet{2009ARA&A..47..481A}. Masses ($M$), luminosities ($L$), and radii ($R$) derived from evolutionary tracks (VLTP or merger tracks) using \Teff and \logg from spectroscopy (see the main text). Alternatively, these quantities (designated as $M_{\rm g}$, $L_{\rm SED}$, and $R_{\rm SED}$) were derived from SED fits along with reddening \ebv, using \emph{Gaia} parallax distances ($d$) from \cite{2021AJ....161..147B}.
     }  } 
\end{center}
\end{table*}

\begin{table}[t]
\tiny
\begin{center}
\caption{
Unidentified photospheric lines. 
  \tablefootmark{a}
}
\label{tab:unidlines} 
\begin{tabular}{ccccc}
\hline 
\hline 
\noalign{\smallskip}
wavelength& \gscs      & \ucacs     \\ 
   (\AA)  & 72\,000\,K & 50\,000\,K \\    \hline
\noalign{\smallskip} 
3505.2    & --         & x  \\
3676.8    & x          & -- \\
3804.2    & --         & x  \\
3819.4    & --         & x  \\
3954.2    & --         & x  \\
4575.6    & x          & -- \\
4579.4    & x          & -- \\
4585.4    & x          & -- \\
4594.6    & x          & -- \\
4619.5    & --         & x  \\
5127.1    & --         & x  \\    
5319.0    & x          & -- \\
5327.4    & x          & -- \\
5440.5    & x          & -- \\
5780.8    & --         & x  \\   
6604.6*   & x          & -- \\
6617.8*   & x          & -- \\
6705.3*   & x          & -- \\
\noalign{\smallskip}
\hline
\end{tabular} 
\tablefoot{\tablefoottext{a}{Effective temperatures of the stars are given below the object names. Emission lines are marked by an asterisk.}}
\end{center}
\end{table}

\subsection{\gsc: A PG1159 star}

 The PG1159 star \gsc (henceforth \gscs) first appeared in the HST Guide Star Catalog (1990). It was initially classified as an OB star by Drilling \& Bergeron (1995) and listed there under the name LSE 41. Later on it was classified as a hot subdwarf by \cite{2022A&A...662A..40C}. Its spectrum is displayed in Fig.\,\ref{fig:gsc}. It shows a line blend made up of \Ion{He}{2} 4686\,\AA\ and several \Ion{C}{4} lines in the range 4640 -- 4700\,\AA. The high-order members of the \Ion{He}{2} Pickering line series point toward a low surface gravity of about \logg $\approx$ 5 in contrast to all other PG1159 stars, which have higher gravities of up to \logg $\approx 8$. The fact that the \ion{He}{ii} 5410\,\AA\ and \ion{C}{iv} 5470\,\AA\ lines have similar strengths means that the helium and carbon abundances are similar. The high C abundance puts \gscs in the class of PG1159 stars. 

We do not see lines from \Ion{He}{1} or \Ion{C}{3}, which allowed us to set a lower limit on the effective temperature. Lines from three oxygen ionization stages are seen (\Ion{O}{4} -- \Ion{O}{6}). The respective ionization balance strictly confines the temperature. We note that the red component of the \Ion{O}{6} doublet at 3811/34\,\AA\ is obviously blended by a relatively strong unidentified line. Mainly from the \Ion{He}{2} lines and the oxygen ionization balance, we find a good compromise at \Teff = 72\,000\,K. Some oxygen lines would fit better at higher and some at lower temperatures. We therefore fixed an error limit of $\pm3000$\,K. The surface gravity is well constrained by the \Ion{He}{2} lines to \logg = $4.8\pm0.2$. For the carbon and oxygen abundance, we find C = $0.15\pm0.05$ and O = $0.23\pm0.05$. Within the error limits, most C and O lines can be fitted. We observe lines from \Ion{N}{4} and \Ion{N}{5}. The nitrogen abundance was found from the \Ion{N}{4} lines, but we assigned it a relatively large error because no simultaneous fit to all lines can be found at the finally adopted value: N = $0.003^{+0.007}_{-0.002}$. For the \Ion{N}{5} lines we find no satisfactory fit. The doublet at 4604/4620\,\AA\ is too weak in the models even if we increase the abundance above N = 0.01. At this high abundance, other lines from this ion appear in the model that are not observed. The \Ion{N}{5} multiplet at 4945\,\AA\ is not matched by any model, probably because of the difficulties with this multiplet. There is no clear detection of silicon in the spectrum, and we derive an upper limit of Si < 0.001. Higher abundances predict \Ion{Si}{4} absorption lines at 3762/73\,\AA\ and at 4212\,\AA, which are not present in the observation. Finally, an upper limit of the hydrogen abundance (H $< 0.001$) was found from the absence of a H$\alpha$ emission line core. The results for \gscs are summarized in Table~\ref{tab:resultsall}, and the best-fit model spectrum is shown in Fig.\,\ref{fig:gsc}.

We note the presence of prominent broad \Ion{C}{4} Rydberg lines similar to the case of \ucacs (see the next subsection). In fact, in the case of \gscs, these features are strongest, for instance, the line at 4230\,\AA. 

\subsection{\ucac: A CO-sdO star}

\ucac (hereafter \ucacs) was classified as a hot subdwarf by \citet{2019A&A...621A..38G}. It was found to be variable based on \emph{Gaia} epoch photometry. The \cite{2022yCat.1355....0G} estimates a period of 74\,min, while \cite{Holl2023} give a period of 59\,min. If the variability were caused by a close companion, then this would imply that CO-sdOs are not the product of a merger. Therefore, time-resolved photometric and spectroscopic follow-up would be highly desirable to confirm the photometric variability and to uncover its origin. We note that based on  \textit{Zwicky} Transient Facility light curves \citep{Bellm+2019, Masci+2019},  neither of the other two CO-sdOs were found to be variable. 

Figure\,\ref{fig:ucac} shows the spectrum of \ucacs. A close inspection reveals that the temperature must be relatively low. There are no \Ion{O}{5} lines present; instead we see a large number of \Ion{O}{3} lines, particularly in the range 3250 -- 3560\,\AA. In addition, we note the presence of many \Ion{C}{3} lines (e.g., near 3890\,\AA) that are not at all visible in the PG1159 star \gscs. And the \Ion{C}{4} 5802/12\,\AA\ doublet is much stronger in absorption than in the PG1159 star. Generally, however, the strengths of the carbon and oxygen lines indicate a chemical composition similar to that of the PG1159 star. We conclude that this object is the third known member of the CO-sdO class. Its spectrum is similar to the other two CO-sdOs presented in \cite{2022MNRAS.511L..66W}, albeit with a far superior resolution and S/N. The temperature of these stars is around 50\,000\,K. Interestingly, we identify the \Ion{C}{4} Rydberg lines in this star too, and they are similar in strength to the two previous CO-sdOs (see for instance the 4230\,\AA\ ($n=7\rightarrow 12$) feature). Although it went unnoticed, one can also recognize this absorption line in their spectra.

The spectrum of \ucacs exhibits lines from \Ion{He}{1}, \Ion{He}{2}, \Ion{C}{3}, \Ion{C}{4}, \Ion{N}{3}, \Ion{N}{4}, \Ion{O}{3}, \Ion{O}{4}, and \Ion{Si}{4}. The lines from neutral helium are rather sensitive to the effective temperature. Also very useful for the temperature determination are the relative strengths of the C, N, and O lines in the ionisation stages III and IV. With increasing \Teff, lines from stage III become weaker while lines from stage IV become stronger. Again, we computed a small set of model atmospheres to determine the atmospheric parameters. We find a best fit at \Teff = $50\,000\pm3000$\,K. Helium lines were used to constrain the surface gravity, and we find \logg = $5.3\pm0.2$. For the hydrogen abundance, an upper limit of H $<0.001$ was determined from the absence of an H$\alpha$ emission line in the red wing of the corresponding \Ion{He}{2} line. For the metal abundances, we measure C $=0.11\pm0.03$, N $= 0.005\pm0.003$, O $=0.03\pm0.02$, and Si $= (3^{+3}_{-1})\times10^{-4}$. The results for \ucacs are summarized in Table~\ref{tab:resultsall}, and the respective best-fit model spectrum is shown in Fig.\,\ref{fig:ucac}.

\subsection{Unidentified photospheric lines}

The spectrum of the O(He) star \gaiafs is rather pure in the sense that it only shows helium lines plus a few lines from nitrogen. The other two objects have a rich spectrum, showing many lines from carbon and oxygen because the abundance of these elements are rather high (C = 0.11 -- 0.15 and O = 0.03 -- 0.23). These two objects also exhibit many lines that we are unable to identify. We suspect that they are mostly unknown lines from highly excited states in \Ion{C}{3}, \Ion{O}{4}, and \Ion{O}{5}. In Table~\ref{tab:unidlines} we list lines that are strongest in at least one of the objects. We note that the Atomic Line List\footnote{\url{https://linelist.pa.uky.edu/atomic/}} \citep{Kentucky} and the NIST Atomic Spectra Database\footnote{\url{https://www.nist.gov/pml/atomic-spectra-database}} were extensively used for our spectral analysis.

\begin{figure}[ht]
 \centering  \includegraphics[width=\columnwidth]{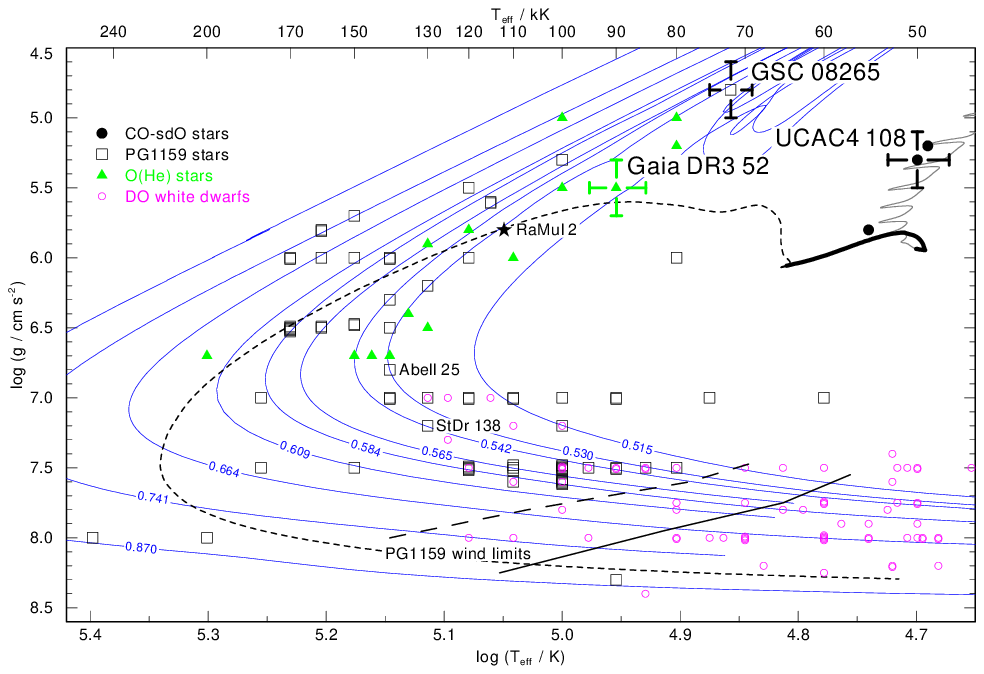}
 \caption{Positions of our program stars in the Kiel diagram (with error bars), together with other hydrogen-deficient stars: CO-sdO stars, PG1159 stars, O(He) stars, and DO WDs. The PG1159 central stars of Abell~25 and StDr~138 and the [WC] central star of RaMul~2 (star symbol) are labeled. Evolutionary tracks (blue lines) for VLTP post-asymptotic giant branch stars labeled with the stellar mass in solar units are from \cite{2006A&A...454..845M}. The post-He-core-burning phase of a 0.8\,\Msol\ remnant of a He+CO WD merger (Z = 0.001, He = 0.32, C = 0.20, O = 0.48) is visualized by the thin gray line, the He-core-burning phase by the thick black line, and the post-He-core-burning phase by the short-dashed black line \citep[from][]{2022MNRAS.511L..60M}. The solid black line indicates the PG1159 wind limit according to \citet{2000A&A...359.1042U}: the mass-loss rate of the radiation-driven wind at this position of the evolutionary tracks becomes so weak that heavy elements are removed from the atmosphere via gravitational settling. Thus, no PG1159 stars should be found at significantly cooler temperatures. The long-dashed line is the wind limit assuming a ten-times-lower mass-loss rate. }
\label{fig:gteff}
\end{figure}

\section{Masses, radii, luminosities, and reddening}
\label{sect:parameters}

Figure\,\ref{fig:gteff} shows the position of the analyzed stars in the Kiel (\Teff -- \logg) diagram. Their masses ($M$) were determined via comparison with evolutionary tracks using linear interpolation or extrapolation. For the O(He) star and the PG1159 star, we found the mass via comparison with the tracks that represent stars that experienced a VLTP \citep{2006A&A...454..845M} and, alternatively by comparing them with binary He WD merger tracks \citep{2012MNRAS.419..452Z}. The merger tracks are not shown here but can be viewed, for instance, in \cite{Werner2022}, their Fig.\,12. For the CO-sdO \ucacs, we used the CO+He WD merger tracks of \cite{2022MNRAS.511L..60M}. The star is close to a track with 0.8\,\Msol; however, this mass estimate is only approximate because tracks of different masses are running relatively close to each other considering the error in the gravity measurement. Using the derived stellar masses (VLTP mass for \gscs and merger mass for the other stars), we calculated stellar radii as $R/R_\odot= [(M/M_\odot)/(g/g_\odot)]^{-1/2}$ and used these radii to compute luminosities as $L/L_\odot = (R/R_\odot)^2(T_\mathrm{eff}/T_{\mathrm{eff},\odot})^4$. 

\begin{figure}
\centering
\includegraphics[width=0.9\columnwidth]{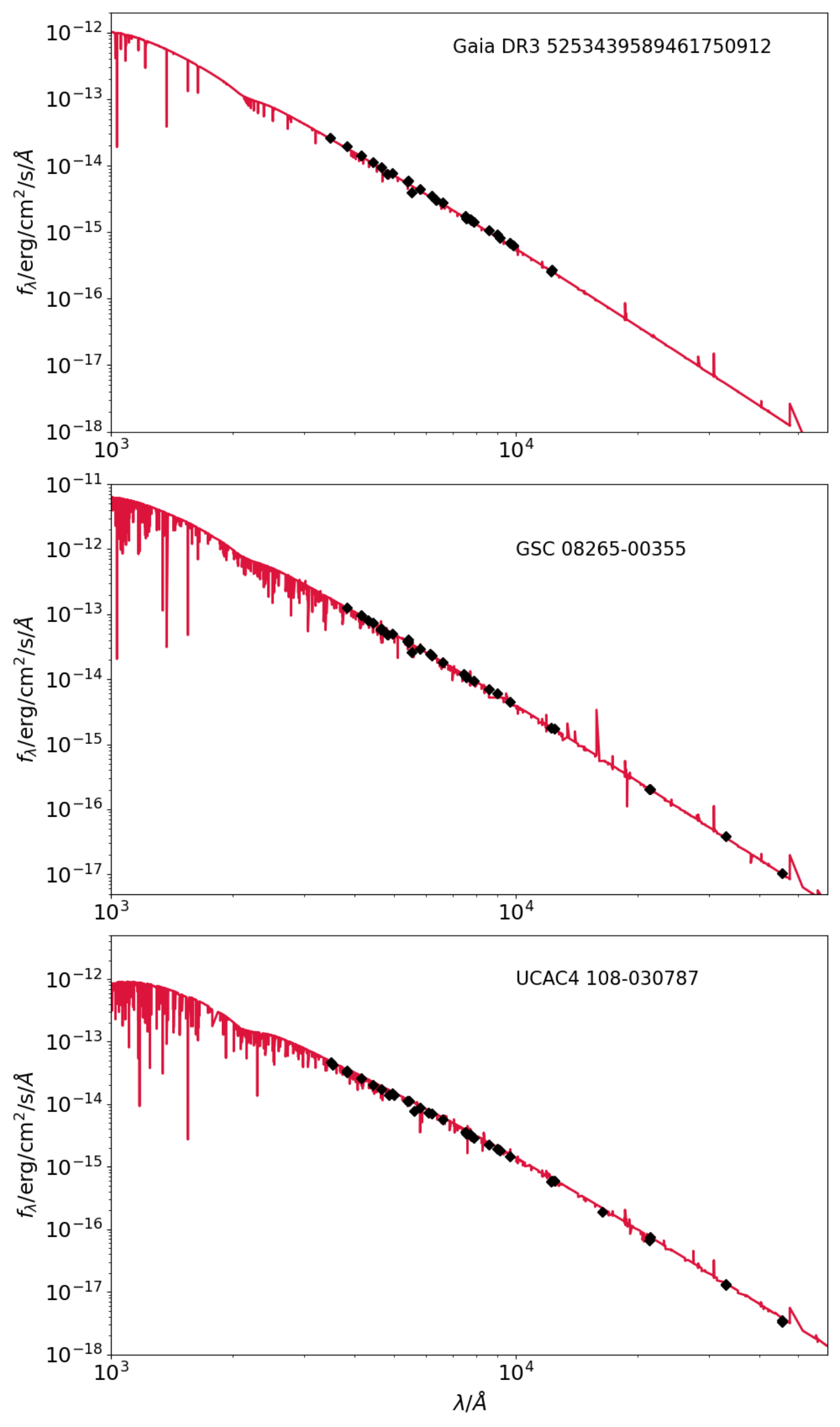}
\caption{SED fits for our three stars. Black dots correspond to observed photometry, and the red lines are our best-fit models corrected for interstellar reddening.}
\label{fig:sed}
\end{figure}

Alternatively, we can use the \emph{Gaia} parallax distances to derive masses, radii, and luminosities, along with the reddening \ebv, from a fit of the model spectrum to the observed spectral energy distribution (SED; see Fig.\,\ref{fig:sed}). The radius $R_{\rm SED}$ was calculated from $f_\lambda = F_\lambda \pi (R_{\rm SED}/d)^2$, where $f_\lambda$ is the observed flux distribution and $F_\lambda$ is the (reddened) astrophysical flux from the model atmosphere. The luminosity was calculated as $L_{\rm SED}/L_\odot = (R_{\rm SED}/R_\odot)^2(T_\mathrm{eff}/T_{\mathrm{eff},\odot})^4$, and the mass as $M_{\rm g}/M_\odot=g/g_\odot (R_{\rm SED}/R_\odot)^2$. 

The results for the stellar parameters are given in Table~\ref{tab:resultsall}. Generally, the masses, radii, and luminosities derived from stellar evolutionary tracks and from the SED fits are consistent.

\section{Summary and discussion}
\label{sect:discussion}

We have analyzed three new hot helium-dominated pre-WDs. Effective temperatures and surface gravities are in the range \Teff = 50\,000 -- 90\,000\,K and \logg = 4.8 -- 5.5. They are all strongly hydrogen-deficient (H < 0.001) and have about six times the solar nitrogen abundance (N = 0.003 -- 0.005).

The first object, \gaiafs, is of spectral type O(He). \gaiafs (\Teff = $90\,000\pm5000$, \logg = $5.5\pm0.2$) has a  CNO abundance pattern that is commonly found in  O(He) stars \citep[e.g.,][]{2014A&A...566A.116R, DeMarco2015, Jeffery2023}. It is enriched in nitrogen and strongly depleted in carbon and oxygen. Since \gaiafs lacks a PN, we can speculate that it formed via the merger of two He-core WDs \citep{2012MNRAS.419..452Z}.

The second object, \gscs, is a PG1159 star (\Teff = $72\,000\pm3000$\,K, \logg = $4.8\pm0.2$). We currently know about 70 PG1159 stars \citep[see][and references therein]{2006PASP..118..183W,2024A&A...686A..29W}, but the new member has unique properties. It is the coolest PG1159 star among its pre-WD cousins (i.e., those that are still evolving toward higher temperatures; see Fig.\,\ref{fig:gteff}) and has the lowest surface gravity. It is believed that Wolf-Rayet-type central stars with PNe transform into PG1159 stars as soon as their radiation-driven winds cease with decreasing luminosity. A recently detected trio of central stars with similar masses (0.53 -- 0.54 \Msol), which represent an evolutionary sequence, confirmed this picture \citep[see Fig.\,\ref{fig:gteff} and][]{2024A&A...686A..29W}. The nucleus of the PN RaMul~2 is a Wolf-Rayet star of intermediate spectral type [WC4--5] with \Teff = 112\,000\,K and \logg = 5.8. The central star of Abell~25 is in a more evolved stage with a higher temperature and gravity (\Teff = 140\,000\,K, \logg = 6.8), while the central star of StDr~138 is even more compact (\logg = 7.2) and already moving downward on the WD cooling sequence (\Teff = 130\,000\,K). 

Our new PG1159 star, \gscs, has the same mass as this trio and represents an earlier evolutionary stage than the [WC] central star of StDr~138. Obviously it is not only the position of the star in the Hertzsprung-Russell diagram that determines whether a significant wind can be driven. Metallicity could play a role, or the abundances of light metals. The values of the carbon and oxygen abundances in \gscs on their own (C = $0.15\pm0.05$ and O = $0.23\pm0.05$) are not unusual for PG1159 stars, but the abundance ratio O/C = $1.5^{+1.3}_{-0.6}$ is exceptional. This ratio is usually less or much less than unity. 

In contrast to the mentioned trio of central stars, \gscs is not known to be associated with a PN. From the nitrogen abundance we can conclude that the star underwent a VLTP and not other variants of this event, namely a LTP or an asymptotic giant branch  final thermal pulse 
\citep[see, e.g.,][]{2006PASP..118..183W,Lawlor2023}; in other words, this event could have happened long after its PN had dispersed. Future spectroscopic surveys might reveal PG1159 stars (with or without associated PNe) in an even earlier post-asymptotic giant branch evolutionary stage than \gscs. 

The third object discussed in this paper, \ucacs, is a hot subdwarf (\Teff = $50\,000\pm3000$\,K, \logg = $5.3\pm0.2$) and only the third known member of the recently identified spectral class CO-sdO \citep{2022MNRAS.511L..66W}. It has been classified as such due to its high abundances of carbon and oxygen (C = $0.11\pm0.03$, O = $0.03\pm0.02$). The first two detected CO-sdOs, however, had significantly higher abundances (C = $0.15\pm0.05$, $0.25\pm0.10$, and O = $0.23\pm0.06$, $0.17^{+0.12}_{-0.07}$). The origin of the CO-sdOs is suspected to be a merger of a low-mass CO WD that was disrupted and accreted by a higher-mass He WD \citep{2022MNRAS.511L..60M}. It is conceivable that the resulting surface abundances of carbon and oxygen depend sensitively on the characteristics of the two merged WDs.

\begin{acknowledgements} 
N.R. is supported by the Deutsche Forschungsgemeinschaft (DFG) through grant RE3915/2-1. The TMAD tool (\url{http://astro.uni-tuebingen.de/~TMAD}) used for this paper was constructed as part of the activities of the German Astrophysical Virtual Observatory. This research has made use of NASA's Astrophysics Data System and the SIMBAD database, operated at CDS, Strasbourg, France. This research has made use of the VizieR catalogue access tool, CDS, Strasbourg, France. This work has made use of data from the European Space Agency (ESA) mission \emph{Gaia}.
\end{acknowledgements}

\bibliographystyle{aa}
\bibliography{aa}

\onecolumn

\begin{appendix}

\onecolumn

\section{Figures}

\FloatBarrier

\begin{landscape}
\addtolength{\textwidth}{6.3cm} 
\addtolength{\evensidemargin}{0cm}
\addtolength{\oddsidemargin}{0cm}
\begin{figure*}[h!]
  \centering  
  \includegraphics[width=0.68\textwidth,angle=-90]{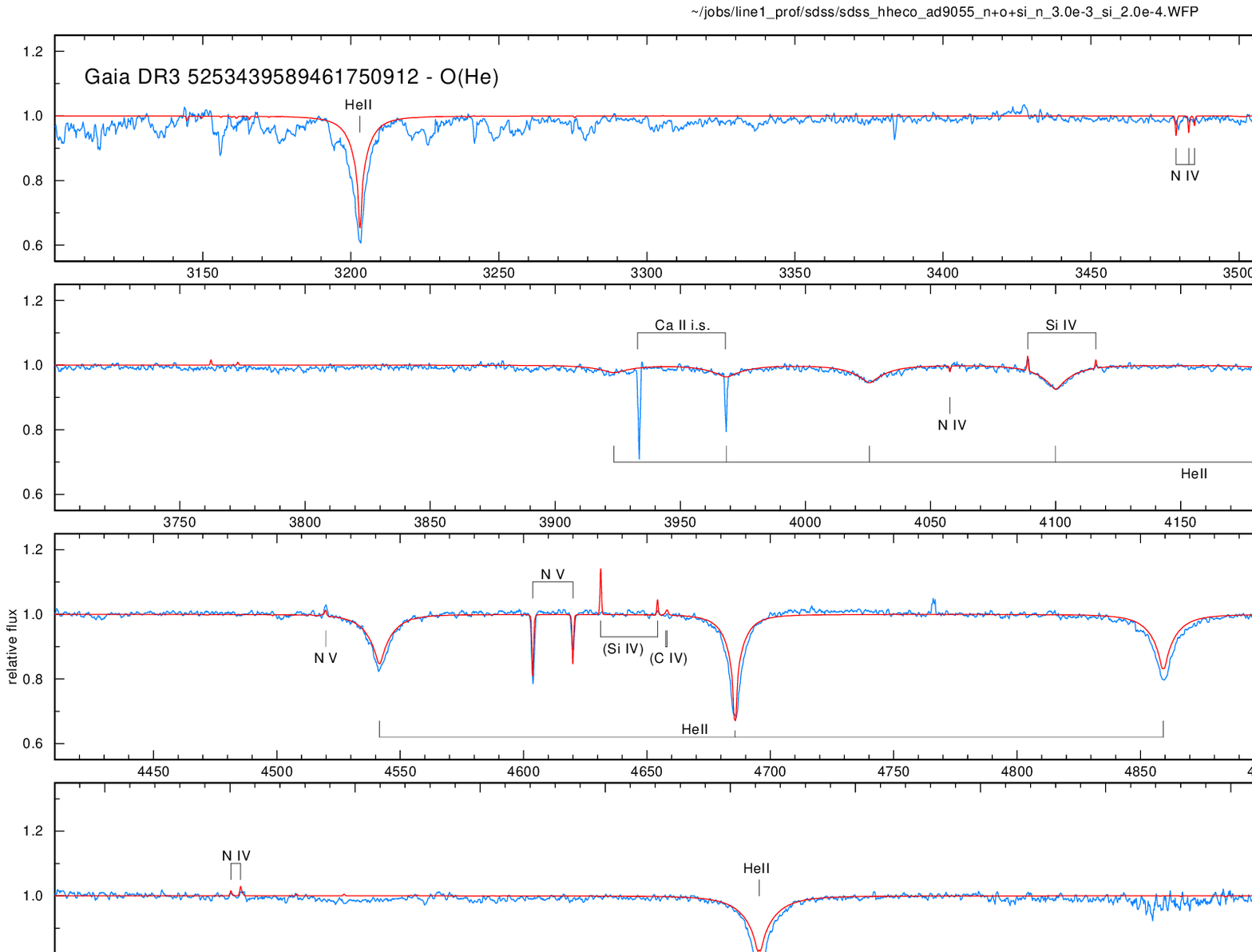}
  \caption{X-Shooter spectrum of the O(He) star \gaiafs. Overplotted is the final model (red) with \Teff = 90\,000\,K, \logg = 5.5, and element abundances as given in Table~\ref{tab:resultsall}. Note the presence of highly excited \ion{C}{iv} lines at locations indicated by horizontal blue and green bars (with principal quantum numbers marked). They are not included in the model except for the 7$\rightarrow$10 transitions at 5470\,\AA. }
\label{fig:gaiaf}
\end{figure*}
\end{landscape}

\begin{landscape}
\addtolength{\textwidth}{6.3cm} 
\addtolength{\evensidemargin}{0cm}
\addtolength{\oddsidemargin}{0cm}
\begin{figure*}[h!]
  \centering  \includegraphics[width=0.68\textwidth,angle=-90]{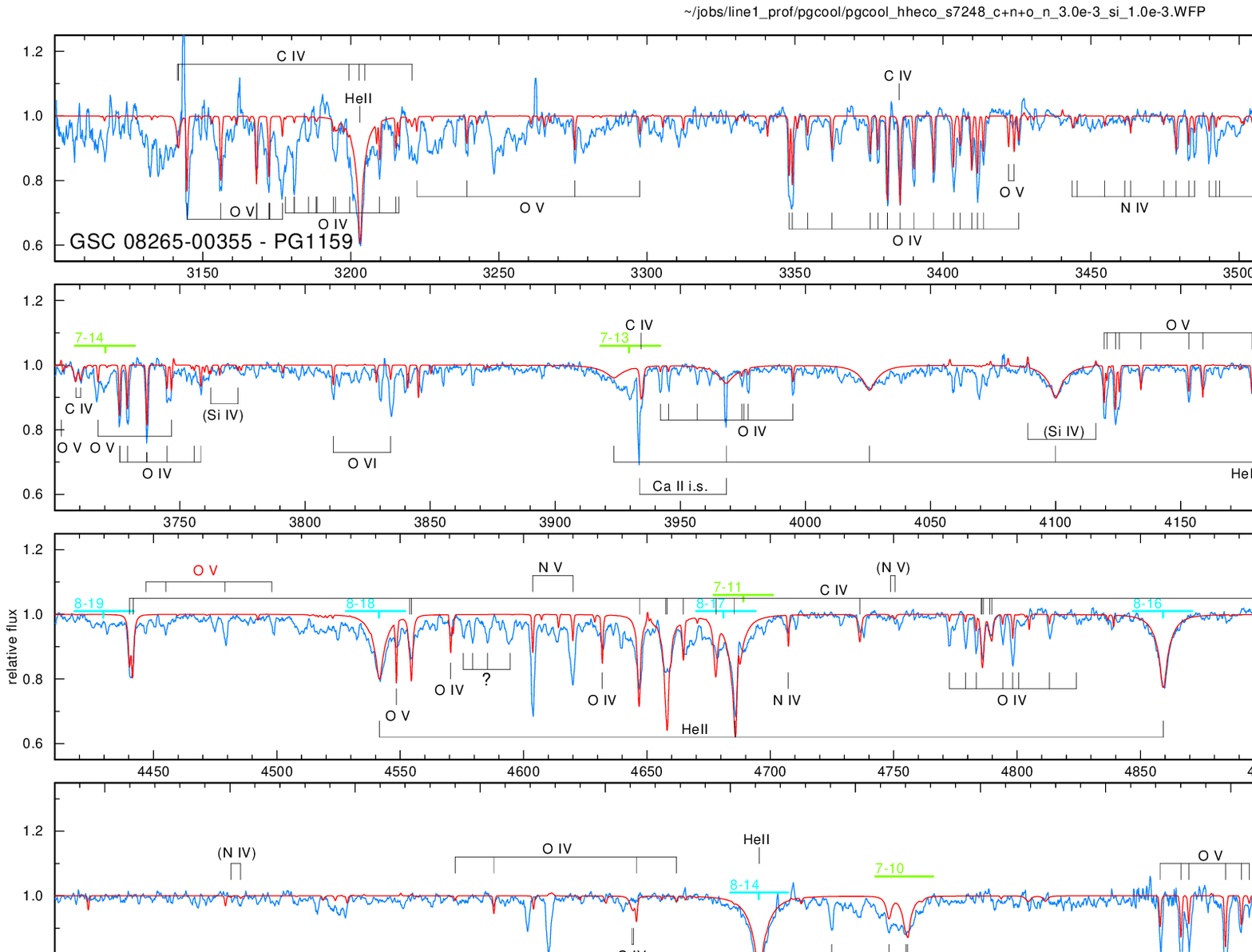}
  \caption{Like Fig.\,\ref{fig:gaiaf}, but for the PG1159 star \gscs. Overplotted is the final model (red) with \Teff = 72\,000\,K, \logg = 4.8, and element abundances as given in Table~\ref{tab:resultsall}. Red identification labels indicate lines that are not included in the model. Note the presence of highly excited \ion{C}{iv} lines at locations indicated by horizontal blue and green bars (with principal quantum numbers marked). They are not included in the model except for the 7$\rightarrow$10 transitions at 5470\,\AA.
  Question marks indicate unidentified photospheric lines.}
\label{fig:gsc}
\end{figure*}
\end{landscape}

\begin{landscape}
\addtolength{\textwidth}{6.3cm} 
\addtolength{\evensidemargin}{0cm}
\addtolength{\oddsidemargin}{0cm}
\begin{figure*}[h!]
  \centering  \includegraphics[width=0.68\textwidth,angle=-90]{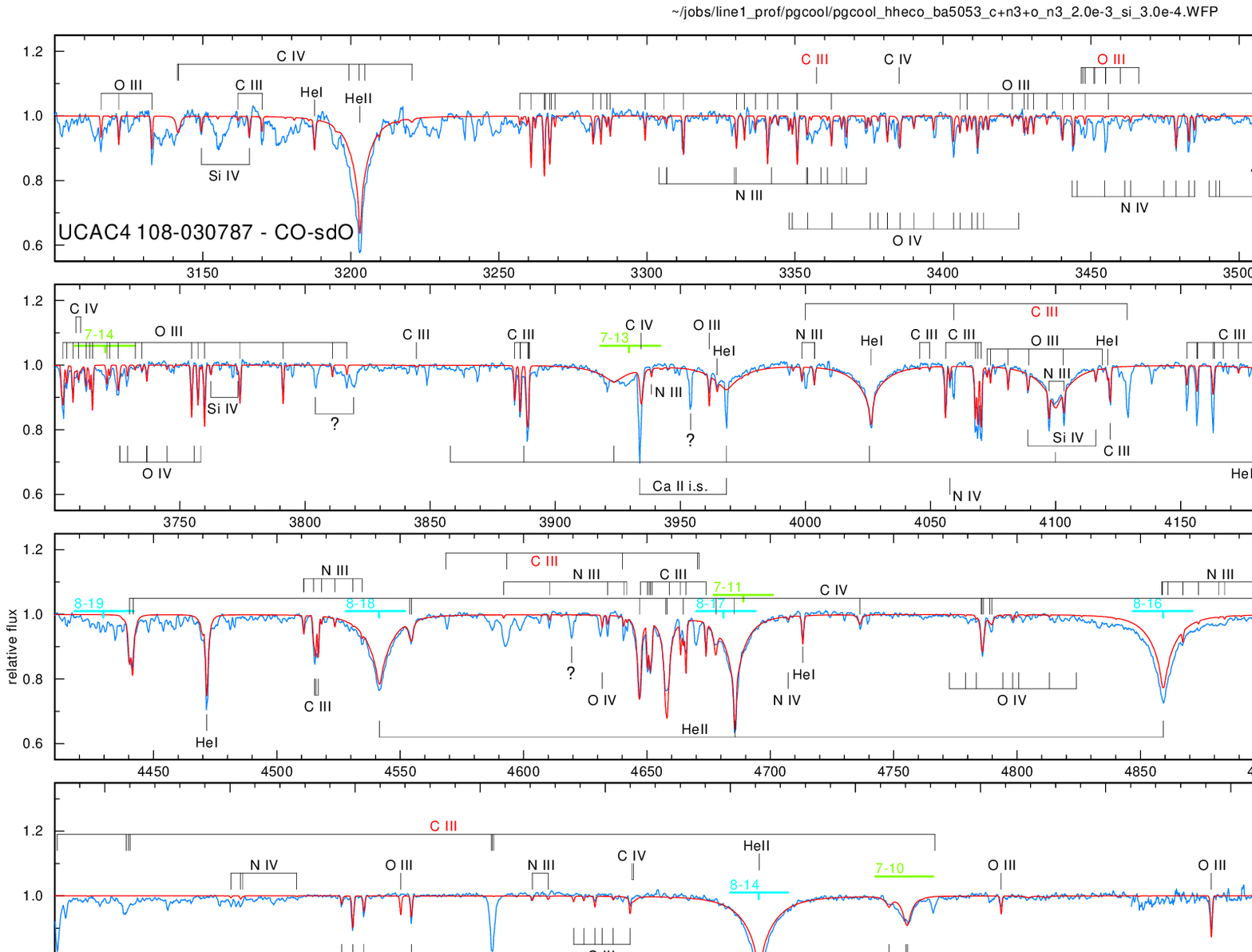}
  \caption{Like Figs.\,\ref{fig:gaiaf} and \ref{fig:gsc}, but for the carbon-oxygen sdO star \ucacs. Overplotted is the final model with \Teff = 50\,000\,K, \logg = 5.3, and its element abundances as given in Table~\ref{tab:resultsall}.}
\label{fig:ucac}
\end{figure*}
\end{landscape}

\end{appendix}
\end{document}